\documentclass[10pt, journal, comsoc]{IEEEtran}
\usepackage{amsmath, amsfonts, amssymb}
\usepackage{algorithmic}
\usepackage{algorithm}
\usepackage{array}
\usepackage{textcomp}
\usepackage{stfloats}
\usepackage{url}
\usepackage[hidelinks]{hyperref}
\usepackage{verbatim}
\usepackage{graphicx, color, xcolor}
\usepackage{cite}
\usepackage{subcaption}

\usepackage{tabularx}
\usepackage{booktabs} 

\begin{document}

\title{Bayesian EM Digital Twins Channel Estimation}
\author{Lorenzo Del Moro$^1$, Francesco Linsalata$^1$, Marouan Mizmizi$^1$, \\ Maurizio Magarini$^1$, Damiano Badini$^2$ and Umberto Spagnolini$^1$
\thanks{$^1$These authors are with Dipartimento di Elettronica, Informazione e Bioingegneria, Politecnico di Milano, Milan, Italy. 
$^2$Damiano Badini is with Huawei Italy Research Center. This work is with Huawei Joint Lab between Politecnico di Milano and Huawei Italy Research Center.}%
\thanks{The work was partially supported by the European Union under the Italian National Recovery and Resilience Plan (NRRP) of NextGenerationEU, partnership on “Telecommunications of the Future” (PE00000001 - program “RESTART”, Structural Project 6GWINET).}
    }



\maketitle

\begin{abstract}

This letter proposes a Bayesian channel estimation method that leverages on the a priori information provided by the Electromagnetic Digital Twin's (EM-DT) representation of the environment. The proposed approach is compared with several conventional techniques in terms of Normalized Mean Square Error (NMSE), spectral efficiency, and number of pilots. Simulations prove more than $10\,$dB gain in NMSE and a spectral efficiency comparable to that of the ideal channel state information, for different signal-to-noise ratio (SNR) values. Additionally, the Bayesian EM-DT-empowered channel estimation enables a remarkable pilot reduction compared to maximum likelihood methods at low SNR.

\end{abstract}


\section{Introduction}
\label{Section 1}


Future sixth-generation (6G) networks will need to support new application scenarios and use cases, imposing high demands on data rates, reliability, efficiency, and latency. To address these challenges, several key enabling technologies are under investigation \cite{6G_requirements1}. Among them, the concept of Network Digital Twin (NDT) is emerging as one of the fundamental paradigm for building future 6G networks \cite{NDT,DT_2}.

The NDT is a virtual replica of a mobile network, capturing the attributes, behaviours, and interactions of both the radio access network (RAN) and core network. Specifically, for a RAN, Electromagnetic Digital Twin (EM-DT) refers to a virtual model of the electromagnetic environment. EM-DT supports both pre- and online analysis based on the physical wireless environment, enabling informed decision-making within communication systems~\cite{zhu2024realtime}. Using detailed 3D maps and ray-tracing software, EM-DT-enhanced networks create accurate, dynamic digital models of the electromagnetic environment in quasi real-time between a base station (BS) and a user equipment's (UE) estimated position~\cite{tan2021integrated}. The EM-DT representation can be exploited for various wireless communication tasks, such as beam selection and blockage management \cite{Beam_man_DT}. In this paper, for the first time, we leverage on the EM-DT as the priori to facilitate channel state information (CSI) acquisition, which is well known to be a critical issue to achieve the optimal performance.

Channel estimation task comprises many approaches.
Pilot-assisted channel estimation is a widely adopted technique to obtain CSI in multi-carrier systems. 
In scenarios with rapidly varying channel conditions, the coherence time decreases significantly, leading to increased overhead and loss in spectral efficiency~\cite{LinsalataBEM}.  
To address this issue,~\cite{Miz/Tag} proposes a low rank model-based channel estimation approach that exploits recurrent vehicle passages to obtain slow time-varying propagation modes of the UE. Recently, research trends in wireless communications are pushing towards data-driven deep learning aided wireless communications as in~\cite{DT_compr_DL}. 
However, there is a lack of extensive research on channel estimation in EM-DT-empowered communications. In this Letter, we propose a Bayesian model-based approach to channel estimation, exploiting the a priori data provided by the EM-DT to obtain the propagation modes of the channel and leveraging a new approach, with the aims of reducing pilot overhead and increasing spectral efficiency.

\textbf{Contributions.} The main novel aspects are:
\begin{itemize}
  \item Definition of a framework to extract a priori information using EM-DT obtained through ray-tracing simulations. 
  \item A solution to address the limitations of rapidly fluctuating faded amplitudes \cite{RT_Sim}, as ray tracing inaccuracies prevent a reliable prediction of ground truth geometric properties. We propose extracting only the slow time-varying components of the channel from the EM-DT and then tracking them by continuously updating the DT over time.
    \item Demonstration of the superior performance of the proposed Bayesian EM-DT-empowered channel estimation compared to state-of-the-art techniques. It is shown a Normalized Mean Squared Error (NMSE) improvement of around $10\,$dB at low signal-to-noise ratio (SNR) values, while numerical results indicate a spectral efficiency close to that of ideal CSI. Furthermore, we examine the effect of pilot reduction on NMSE and compare it with the least square (LS) channel estimation approach.
\end{itemize}

\textbf{Organization.} The rest of the paper is organized as follows. Section~\ref{System Model} introduces the system model. In Sec. \ref{DT_met}, we propose our Bayesian EM-DT-empowered channel estimation method, while in Sec. \ref{simulations} numerical results are presented. Section~\ref{Conclusion} concludes the work and suggests future exploration.

\textbf{Notation.} $(\cdot)^*$, $(\cdot)^T$, $(\cdot)^H$, and $(\cdot)^{\perp}$ denote the complex conjugate, matrix transpose, Hermitian transpose, and the complementary matrix. Bold lowercase and uppercase letters represent vectors and matrices, respectively. The Kronecker product is $\otimes$, $\mathrm{diag}(\mathbf{v})$ is a diagonal matrix with vector $\mathbf{v}$ on main diagonal, and $\mathrm{vec}\{\mathbf{A}\}$ is the stacking operator, such that $\mathrm{vec}\{\mathbf{ABC}\}=(\mathbf{C}^T\otimes\mathbf{A})\mathrm{vec}\{\mathbf{B}\}$. Additionally, $\text{Tr}\{\mathbf{A}\}$ is the trace of the square matrix $\mathbf{A}$, $\mathrm{E}\{\cdot\}$ denotes the expectation, and $\mathbf{I}_N$ is the $N \times N$ identity matrix.

\section{System model}
\label{System Model}


\subsection{Transmitted and Received Signal} \label{OFDM}
The reference scenario consists of a single antenna UE that transmits an orthogonal frequency-division multiplexing (OFDM) signal to a BS equipped with $N_{rx}$ antennas. Let $\textbf{x}[m]$$\,\in\,$$\mathbb{C}^{N \times 1}$ denote the vector of complex baseband symbols generated by the UE in the $m$th OFDM symbol. Without loss of generality, we assume i.i.d. symbols, so that $\mathbb{E}\{\textbf{x}[m]\textbf{x}^H[m]\}$$\,=\,$$\sigma^2_x\mathbf{I}_N  $. The duration of the transmitted OFDM symbol is $T $$\,=\,$$ (N + N_{CP})T_S$, where $T_S $$\,=\,$$ \frac{1}{B}$ is the sampling interval given the overall system bandwidth $B$ and $N_{CP}$ is the length of the cyclic prefix (CP). 

By assuming a CP long enough to prevent the intersymbol interference, the complex equivalent baseband received signal in the $m$th OFDM symbol can be expressed in matrix form as
\begin{equation}
\label{OFDM}
    \textbf{Y}[m] = \textbf{H}[m] \mathrm{diag}(\textbf{x}[m])+ \textbf{W}[m],
\end{equation}
where $\mathbf{H}[m]$$\,\in\,$$ \mathbb{C}^{N_{rx}\times N}$ is the equivalent baseband discrete-frequency channel response between the UE and the $N_{rx}$ antennas of the BS and $\textbf{W}[m] $$\,\in\,$$ \mathbb{C}^{N_{rx}\times N}$ is a matrix with i.i.d. complex Gaussian elements with zero mean and variance $\sigma^2_w$. 

In the context of an OFDM system, pilot-assisted channel estimation is required to estimate the channel and implement coherent demodulation. To this end, pilot symbols are uniformly allocated over ${N_p}$ subcarriers and used for channel estimation, followed by conventional linear interpolation \cite{Miz/Tag}. The estimated channel is used to compute a linear spatial combiner $\mathbf{S}[m] $$\,\in\,$$ \mathbb{C}^{N \times N_{rx}}$ weighting the received data samples as
\begin{equation}
    \mathbf{Z}[m] = \mathbf{S}[m]\textbf{Y}[m] = \mathbf{S}[m]\left(\mathbf{H}[m] \mathrm{diag}(\mathbf{x}[m]) + \mathbf{N}[m]\right).
\end{equation}
The soft estimated transmitted symbols $\hat{\mathbf{x}}[m]$ are finally obtained taking the diagonal elements of $\mathbf{Z}[m]$.

\subsection{Channel Model}
\label{Channel}
The model of the time-varying multi-path channel with $L$ components is hereafter introduced. The equivalent discrete-time faded channel between the UE and the BS associated with the $m$th OFDM symbol can be written in vector form as
\begin{equation}
\label{channel_vec_time}
    \mathbf{h}[m;\nu] = \sum_{l=1}^{L} c_{l}[m]\boldsymbol{a}(\theta_{l}[m],\varphi_{l}[m])g\left(\nu-\frac{\tau_{l}[m]}{T_s}\right),
\end{equation}
where $\nu=0,\ldots,N-1$ is the temporal index, $\tau_{l}[m]$ represents the delay for the $l$th path in the $m$th OFDM symbol, $c_{l}[m]$ the faded amplitude and  $\theta_{l}[m]$, and $\varphi_{l}[m]$ are the elevation and azimuth angle of arrival (AoA), respectively, while $g(\cdot)$ is the impulse response given by the cascade of transmitting and receiving pulse shaping filters. Moreover, for a uniform linear array, the response associated with the $l$th path is $\boldsymbol{a}(\theta_{l}[m],\varphi_{l}[m])=[e^{j\frac{2\pi}{\lambda}\mathbf{u}^T_1\mathbf{v}_l},...,e^{j\frac{2\pi}{\lambda}\mathbf{u}^T_{N_{rx}}\mathbf{v}_l}]^T$, where $\lambda$ is the 
wavelength, $\mathbf{u}_i$, $i=1,\ldots,N_{rx}$ is the position of the $i$th array element, and 
\begin{equation}
    \mathbf{v}_l\hspace{-.1cm}=\hspace{-.1cm}\begin{bmatrix}\hspace{-.03cm}
\cos(\hspace{-.03cm}\varphi_l[\hspace{-.03cm}m\hspace{-.03cm}]\hspace{-.03cm})\cos(\hspace{-.03cm}\theta_l[\hspace{-.03cm}m\hspace{-.03cm}]\hspace{-.03cm}), 
\sin(\hspace{-.03cm}\varphi_l[\hspace{-.03cm}m\hspace{-.03cm}]\hspace{-.03cm})\cos(\hspace{-.03cm}\theta_l[\hspace{-.03cm}m\hspace{-.03cm}]\hspace{-.03cm}),
\sin(\theta_l[m])
    \hspace{-.2cm}\end{bmatrix}^T.\hspace{-.18cm}
\end{equation} 
By computing the DFT of the $N$ consecutive vector faded channels in~\eqref{channel_vec_time}, we obtain the following space-frequency domain matrix representation of the channel
\begin{equation}
\label{channel_split}
    \textbf{H}[m] = \mathbf{A}(\boldsymbol{\theta}[m],\boldsymbol{\varphi}[m])\textbf{C}[m]\mathbf{K}(\boldsymbol{\tau}[m])^T,
\end{equation}
where 
\begin{itemize}
    \item $\mathbf{A}(\boldsymbol{\theta}[m],\boldsymbol{\varphi}[m]) \hspace{-.1cm}= \hspace{-.1cm}\left[\boldsymbol{a}(\theta_{1}\hspace{-.05cm}[m],\hspace{-.05cm}\varphi_{1}\hspace{-.05cm}[m]),\ldots,\boldsymbol{a}(\theta_{L}\hspace{-.05cm}[m],\hspace{-.05cm}\varphi_{L}\hspace{-.05cm}[m])\right]$ is the matrix obtained from the composition of the array responses for the $L$ paths;
    \item $\mathbf{C}[m]=\mathrm{diag}(\mathbf{c}[m])$, with $\mathbf{c}[m] =\left[c_1[m],...,c_L[m]\right]^T$;
    \item $\mathbf{K}(\boldsymbol{\tau}[m])\in \mathbb{C}^{N\times L}$ is the frequency response matrix given by $\mathbf{F}\mathbf{G}(\mathbf{\tau}[m])$, where $\mathbf{F}$ is the $N$-point DFT matrix and $\mathbf{G}(\mathbf{\tau}[m])=\left[\mathbf{g}(\tau_1[m]),...,\mathbf{g}(\tau_L[m])\right]$, with $\mathbf{g}(\tau_l[m])=\left[g(-\frac{\tau_l[m]}{T_s}),...,g(N-1-\frac{\tau_l[m]}{T_s})\right]^T$.
\end{itemize} 
In \eqref{channel_split}, the columns of $\mathbf{A}(\boldsymbol{\theta}[m],\boldsymbol{\varphi}[m])$ and of $\mathbf{K}(\boldsymbol{\tau}[m])$ span the spatial and the temporal subspace, respectively, as demonstrated in~\cite{Simeone_basis}. The spatial and the temporal subspaces have a dimension equal to the number of resolvable paths in the angular and the temporal domain given the array length and the system resolution. The dimension of the spatial subspace is therefore  $r_S$$\,=\,$$\mathrm{rank}(\mathbf{A}(\boldsymbol{\theta}[m],\boldsymbol{\varphi}[m]))$$\,\leq\,$$\mathrm{min}(N_{rx},L)$, while the dimension of the temporal subspace is $r_T $$\,=\,$$ \mathrm{rank}(\mathbf{K}(\boldsymbol{\tau}[m]))$$\,\leq\,$$ \mathrm{min}(N,L)$. The model defined in \eqref{channel_split} is the key to separate the slow time-varying components of the channel $\mathbf{A}(\boldsymbol{\theta}[m],\boldsymbol{\varphi}[m])$ and $\mathbf{K}(\boldsymbol{\tau}[m])$ from the fast time-varying components $\textbf{C}[m]$. Specifically, $\mathbf{C}[m]$ is assumed to be characterized by independent realizations across OFDM symbols, i.e. $\mathrm{E}[\mathbf{c}[m]\mathbf{c}[m]^T]$$\,=\,$$\mathrm{diag}(\alpha^2_1[m],...,\alpha^2_L[m])$, where $\alpha_l[m]$ is the amplitude of the the $l$th path. The slow time variability of the channel in a period $T_B$ is related to the relative speed between UE and BS \cite{Simeone_basis}. Coherently with this consideration, hereafter, the overall channel relatively to the $m$th OFDM symbol in a period $T_B$ is
\begin{equation}
\label{channel_const}
    \textbf{H}[m] = \mathbf{A}(\boldsymbol{\theta},\boldsymbol{\varphi})\textbf{C}[m]\mathbf{K}(\boldsymbol{\tau})^T.
    \end{equation} 

\section{Proposed Bayesian EM-DT-empowered channel Estimation Method}\label{DT_met}


 \subsection{Electromagnetic Digital Twin Framework}
 
An EM-DT consists in a fully integrated virtual model mirroring the components and behavior of the EM environment \cite{zhu2024realtime}. The EM-DT and its physical counterpart continuously exchange information in real-time, ensuring that any change in one is reflected in the other.

\begin{figure}[!t]
\centerline{\includegraphics[scale=0.30]{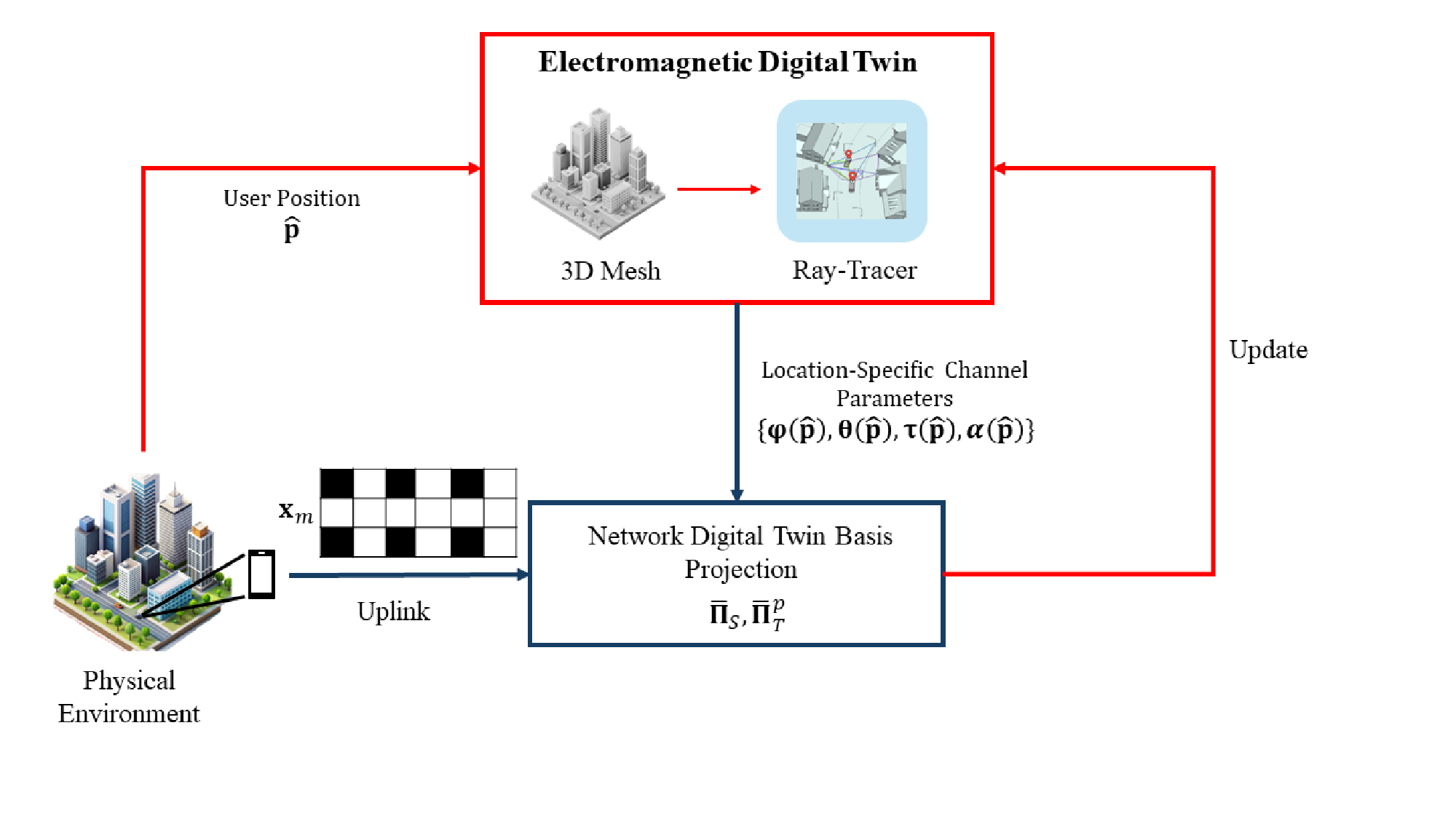}}
\caption{Reference EM-DT framework.}
\label{figEM-DT}
\vspace{-.3cm}
\end{figure}

In this work, we introduce an EM-DT framework to enable environment-aware channel estimation. 
With reference to Fig. \ref{figEM-DT}, the EM-DT is continuously updated with the goal of maintaining an accurate and coherent information on the UE predicted location $\hat{\mathbf{p}}$ at future instants and the physical EM environment. 
At each update, the digital replica of the environment is refreshed, and a ray-tracing simulation is performed to generate the simulated location-specific channel parameters $\{\mathbf{\varphi}(\hat{\mathbf{p}}),\mathbf{\theta}(\hat{\mathbf{p}}),\mathbf{\tau}(\hat{\mathbf{p}}),\mathbf{\alpha}(\hat{\mathbf{p}})\}$, i.e. AoAs, delays and multipath amplitudes based on the estimated UE's position. Since the ray-tracing simulation requires a computational time $T_{DT}$ \cite{zhu2024realtime}, it is important to have the predicted information to be synchronized with $T_{DT}$ so that the location-specific parameters can be used to characterize the real-time spatial and temporal subspaces. The location-specific parameters are then provided to the physical system to improve channel estimation during uplink communication. 
At the same time, the physical system gathers information, which helps the prediction of the UE's state at future instants. Finally, the estimated position is fed back into the EM-DT, completing the cycle of updates and synchronization between the physical world and its digital counterpart. An EM-DT spatial and temporal subspaces description, mainly defined by the geometry of the environment, i.e. delay, angles, and multipath amplitudes, is restricted to only $\bar{L}$ paths \cite{zhu2024realtime}, where $\bar{L} $$\,<\,$$ L$, due to the computational constraints required for real-time characterization of propagation.

We remark that the framework here described allows us for the creation of realistic datasets that integrate synchronized positional and wireless channel data. 

\subsection{Bayesian Basis Projection}

The EM-DT environment extracted from the proposed framework is used as a priori information to characterize the spatial and temporal channel subspaces related to the first $\bar{L}$ paths. Specifically, the a priori information consists of the subspaces obtained from the EM-DT, which are exploited for Bayesian projection in reduced rank channel estimation. Hereafter, we denote the variables limited to $\bar{L}$ paths by a proper overline. The partial channel according to employed evenly spaced pilots, limited to the $\bar{L}$ paths, can be modelled as
\begin{equation}
\label{channel_const}
    \bar{\textbf{H}}^p[m] = \bar{\mathbf{A}}(\boldsymbol{\theta},\boldsymbol{\varphi})\bar{\textbf{C}}[m]\bar{\mathbf{K}}^p(\boldsymbol{\tau})^T.
    \end{equation} 
We remark that while $\bar{\textbf{C}}[m]$ is unknown to the EM-DT, the slow time-varying matrices $\{\bar{\textbf{A}}(\boldsymbol{\theta},\boldsymbol{\varphi}),\bar{\textbf{K}}^p(\boldsymbol{\tau})\}$ spanning the spatial and temporal subspaces are obtained as a priori from the EM-DT. Let $\bar{\mathbf{U}}_S $$\,\in\,$$ \mathbb{C}^{N_{rx} \times\bar{r}_S }$ and $\bar{\mathbf{U}}^p_T$$\,\in\,$$ \mathbb{C}^{N_p\times \bar{r}_T}$ be the corresponding orthonormal basis (propagation modes), such that $\mathrm{span}(\bar{\textbf{A}}(\boldsymbol{\theta},\boldsymbol{\varphi}))\equiv \mathrm{span}(\bar{\mathbf{U}}_S)$ and $\mathrm{span}(\bar{\textbf{K}}^p(\boldsymbol{\tau}))\equiv \mathrm{span}(\bar{\mathbf{U}}^p_T)$. Also, let $\bar{r}_S = \mathrm{rank}(\bar{\textbf{A}}(\boldsymbol{\theta},\boldsymbol{\varphi}))$ and $\bar{r}_T = \mathrm{rank}(\bar{\textbf{K}}^p(\boldsymbol{\tau}))$ be the dimensions of the spatial and temporal subspaces obtained from the EM-DT, respectively. 
It is worth noting that differently from \cite{Miz/Tag}, where a non-negligible latency is introduced by the estimation of the subspace basis from the spatial and temporal sampled channel covariance matrices, the EM-DT allows us to pre-compute and maintain in a period $T_B$ the a priori basis, by exploiting the knowledge on a limited subset of delays and AoAs. These bases $\{\bar{\mathbf{U}}_S,\bar{\mathbf{U}}^p_T\}$ are obtained from the singular value decomposition of $\{\bar{\textbf{A}}(\boldsymbol{\theta},\boldsymbol{\varphi}),\bar{\textbf{K}}^p(\boldsymbol{\tau})\}$.
The projectors related to the spatial and temporal propagation modes are
\begin{equation}
    \label{Projectors}
\begin{aligned}
  \bar{\mathbf{\Pi}}_S=\bar{\mathbf{U}}_S\bar{\mathbf{U}}^{H}_S ,\,\,
    \bar{\mathbf{\Pi}}^p_T=\bar{\mathbf{U}}^{p*}_T\bar{\mathbf{U}}^{pT}_T.
\end{aligned}
\end{equation}
As a consequence, the only operation that must be performed in real-time consists of projecting the LS estimate on the a priori propagation modes through the projectors. Accordingly, this makes the proposed Bayesian method compliant with the current standard architecture and with low computational cost behaving as modal filter. Afterwards, the EM-DT-empowered channel estimation becomes
\begin{equation}
\label{DT_proj}
  \hat{\bar{\textbf{H}}}^p[m] = \bar{\mathbf{\Pi}}_S \textbf{Y}^p[m] \mathrm{diag}(\mathbf{x}^p[m])^{-1}\bar{\mathbf{\Pi}}^p_T ,
\end{equation}
where $\textbf{Y}^p[m] \mathrm{diag}(\mathbf{x}^p[m])^{-1}$ is the LS channel estimation over pilots. Equation~\eqref{DT_proj} highlights that our proposed method is equivalent to a Bayesian channel estimation exploiting the partial a priori propagation modes given by the EM-DT according to the UEs position.
\begin{figure*}[!t]
    \begin{equation}
    \label{vec_err}
    \Delta\mathbf{h}^p[m] = \mathbf{h}^p[m] - (\bar{\mathbf{\Pi}}^{p\:T}_T \otimes \bar{\mathbf{\Pi}}_S) \mathbf{h}^p[m] 
    - (\bar{\mathbf{\Pi}}^{p\:T}_T \otimes \bar{\mathbf{\Pi}}_S) \, \text{vec}\left \{\text{diag}(\mathbf{x}^p[m])^{-1} \mathbf{W}^p[m])\right \}.
    \end{equation}
    \vspace{-0.4cm}
    \begin{equation}
    \label{NMSE}
  \text{NMSE} = \sigma_{\bar{L}}^2 + \sigma_{Qw}^2 =\frac{\text{Tr}\{\mathrm{E}\{\Delta\mathbf{h}^p[m]\Delta\mathbf{h}^p[m]^H\}\}}{\text{Tr}\{\mathbf{R}^p\}} = \frac{\text{Tr}\{\:(\mathbf{Q}^\perp\mathbf{R}^p{\mathbf{Q}^\perp}^H\:)\}}{\text{Tr}\{\mathbf{R}^p\}}
    + \frac{\sigma^2_w\:\text{Tr}\{\mathbf{Q}\mathbf{Q}^H\}}{\sigma^2_x\text{Tr}\{\mathbf{R}^p\} },
    \end{equation}
\end{figure*}

In order to derive the NMSE across pilot subcarriers for the proposed method, we assume the perfect accuracy of the projectors and rewrite~\eqref{DT_proj} by substituting the input-output relation described in \eqref{OFDM} over the pilot subcarriers as
\begin{equation}
\label{HDT_ls}
    \hat{\bar{\textbf{H}}}^p[m] = \bar{\mathbf{\Pi}}_S (\textbf{H}^p[m]+ \mathrm{diag}(\mathbf{x}^p[m])^{-1}\mathbf{W}^p[m])\bar{\mathbf{\Pi}}^p_T .
\end{equation}
The NMSE on the subcarriers is obtained from the covariance of the vectorized error $\Delta\mathbf{h}^p=\mathrm{vec}\{\mathbf{H}^p[m]-\hat{\bar{\mathbf{H}}}^p[m]\}$ that, by exploiting the property of Kronecker operator, is evaluated as given in \eqref{vec_err}.
The NMSE of $\Delta\mathbf{h}^p[m]$ can be obtained as in \eqref{NMSE}, where $\mathbf{R}^p = \mathrm{E}\{\mathbf{h}^p[m]\mathbf{h}^{pH}[m]\}$ is the covariance of the vectorized channel over the pilot subcarriers and $\mathbf{Q} = \bar{\mathbf{\Pi}}^{p\:T}_T \otimes \bar{\mathbf{\Pi}}_S$ is the projection matrix obtained as the Kronecker product between the temporal and spatial projection matrices provided by the EM-DT.
From \eqref{NMSE}, it is evident that the NMSE consists of two distinct components. The first term $\sigma_{\bar{L}}^2$ accounts for the limited number of paths computed by the EM-DT and sets a floor on the NMSE. However, if the known paths are the most significant, the contribution of the unknown ones can be neglected, as shown in Sec.~\ref{simulations}. The second term $\sigma_{Q_W}^2$, which is due to the noise $\mathbf{W}[m]$, can be simplified as 
%
\begin{equation}
\sigma_{Qw}^2 = \frac{\sigma^2_w\:\text{Tr}\{\mathbf{Q}\mathbf{Q}^H\}}{\sigma^2_x\text{Tr}\{\mathbf{R}^p\} }=\frac{\bar{r}_S\bar{r}_T}{N_{rx}N_p\mathrm{SNR}}, 
\end{equation}
where the property $\text{Tr}\{\mathbf{Q}\mathbf{Q}^H\} =\text{Tr}\{\mathbf{Q}
 \}=\bar{r}_Q=\bar{r}_S\bar{r}_T$ has been applied and where the signal-to-noise ratio (SNR) is
\begin{equation}
    \mathrm{SNR}=\frac{\sigma^2_{x}\beta}{\sigma^2_w},
\end{equation}
with $\beta $$\, =\,$$ \frac{1}{N^pN_{rx}}\text{Tr}\{\mathbf{R}^p\}$ defining the average per antenna and per subcarrier channel gain.
\begin{figure}[!t]
\centerline{\includegraphics[scale=0.44]{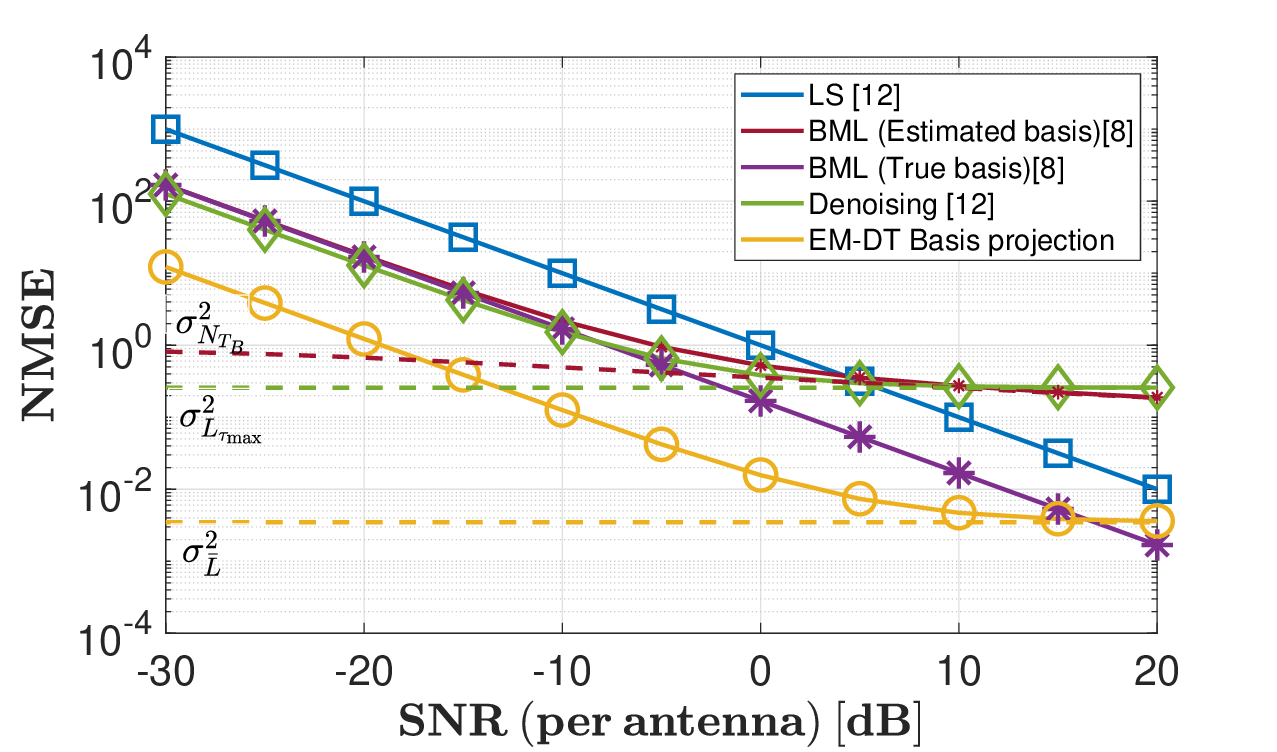}}
\caption{NMSE comparison with $N_p = 32$ among state-of-the-art approaches and our proposed method. Solid lines are the analytical results, while markers denote the simulated results. Dashed lines are the normalized error variance thresholds $\sigma^2_{N_{T_B}}$, $\sigma^2_{L_{\tau_\mathrm{max}}}$and $\sigma^2_{\bar{L}}$ for the BML, the denoising and the proposed method, respectively.}
\label{figNMSE}
\end{figure}
\begin{figure}[!t]
\centerline{\includegraphics[scale=0.44]{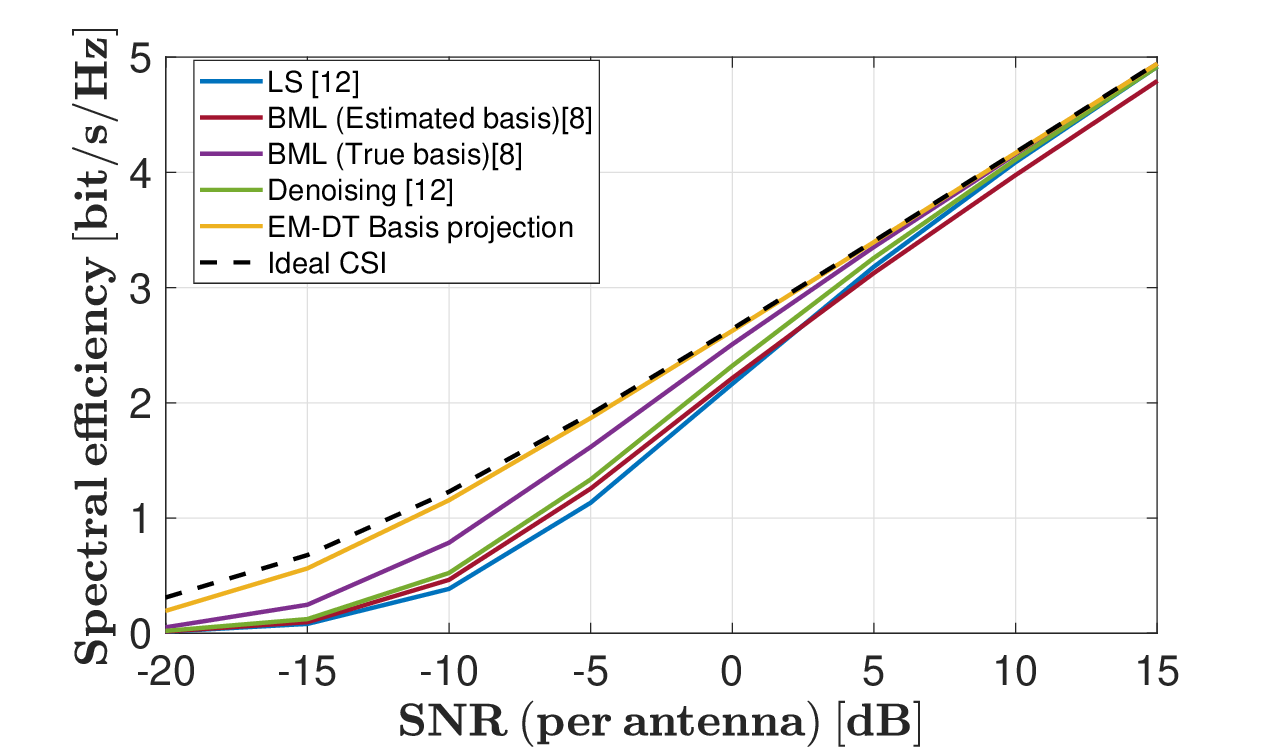}}
\caption{Spectral efficiency comparison when $N_p = 32$ among state-of-the-art approaches and our proposed method.}
\label{figSE}
\end{figure}

Note that, as discussed in \cite{zhu2024realtime}, the proposed method lays on a trade-off between the computational time required to the EM-DT to provide an updated information on the spatial and temporal subspaces, and the number of paths generated from the ray-tracing simulation. In fact, as the number of generated paths increases, the correspondent computational time grows and vice versa. Given the dynamicity of the scenario, which imposes constraints on computational time, the generated paths are restricted to a subset of $\bar{L}$ paths to maintain a balance between efficiency and accuracy.

\section{Numerical results}
\label{simulations}
Consistently with the system model in Sec. \ref{System Model}, we set $N $$\,=\,$$ 64$ and $N_{rx}$$\,=\,$$64$. We allocate $N_p$ pilot symbols every OFDM symbol. We consider numerology $\mu $$\,=\,$$ 5$ of 5G NR frame, corresponding to a subcarrier spacing of $\Delta f $$\,=\,$$ 480\:\mathrm{KHz}$ and to an overall bandwidth $B $$\,=\,$$ 30.72\:\mathrm{MHz}$. The simulated benchmark channel is generated by means of a commercial ray-tracer with $L$$\,=\,$$25$ paths with a carrier frequency $f_c $$\,=\,$$ 28\:\mathrm{GHz}$ \cite{zhu2024realtime}. 
A denoising method proposed in \cite{Den} is evaluated here, which consists in pruning the channel impulse response (CIR) based on a maximum delay spread, $\tau_{max}$. In this simulation, $\tau_{max}=0.5\:\mathrm{\mu s}$ is adopted. Concerning our proposed method, we suppose to have $\bar{L} $$\,=\,$$ 5$ paths, with perfect accuracy. Additionally, we also investigate a batch maximum likelihood (BML) approach, which involves estimating the subspaces basis from the spatial and temporal covariance matrix of the channel and then projecting the LS estimates onto the aforementioned basis. 

Figure \ref{figNMSE} depicts the NMSE for different channel estimation methods. Conversely to the LS approach, which suffers poor performance in the low SNR region, the BML and the denoising approaches guarantee better results improving the performances of roughly $10\,$dB  with respect to the LS, while our Bayesian basis projection overperforms the LS of roughly $20\,$dB. In addition to the latency introduced to estimate the basis (i.e. propagation modes) of the spatial and temporal subspaces, the BML is sensitive to the number of LS estimates $N_{T_B}$ used to derive the spatial and temporal sampled covariance matrix. This limited batch of symbols is more evident at high SNR, as clearly shown in Fig. \ref{figNMSE}, and set a thresholds $\sigma^2_{N_{T_B}}$ on the NMSE over the pilot subcarriers. Moreover, also the denoising approach has a threshold $\sigma^2_{L_{\tau_\mathrm{max}}}$due to the pruned CIR, which in this setup is even higher with respect to the one of the BML for high SNR values, whereas the lowest threshold is that of the EM-DT basis projection $\sigma^2_{\bar{L}}$ which guarantees high performance even at high SNR.

Figure \ref{figSE} shows the ''genie-aided'' upper bound to spectral efficiency, which is evaluated by applying a maximum ratio combiner at the receiver side using channel estimates, but in the final decision step a perfect channel knowledge is assumed. The results highlight the better performance of the proposed method in the low SNR region. In fact, due to the Bayesian projection our method approaches the ideal CSI condition even when the SNR is low, while at high SNR the limited number of known paths does not have impact since $\sigma^2_{\bar{L}}$ can be considered negligible.
\begin{figure}[!t]
\centerline{\includegraphics[scale=0.44]{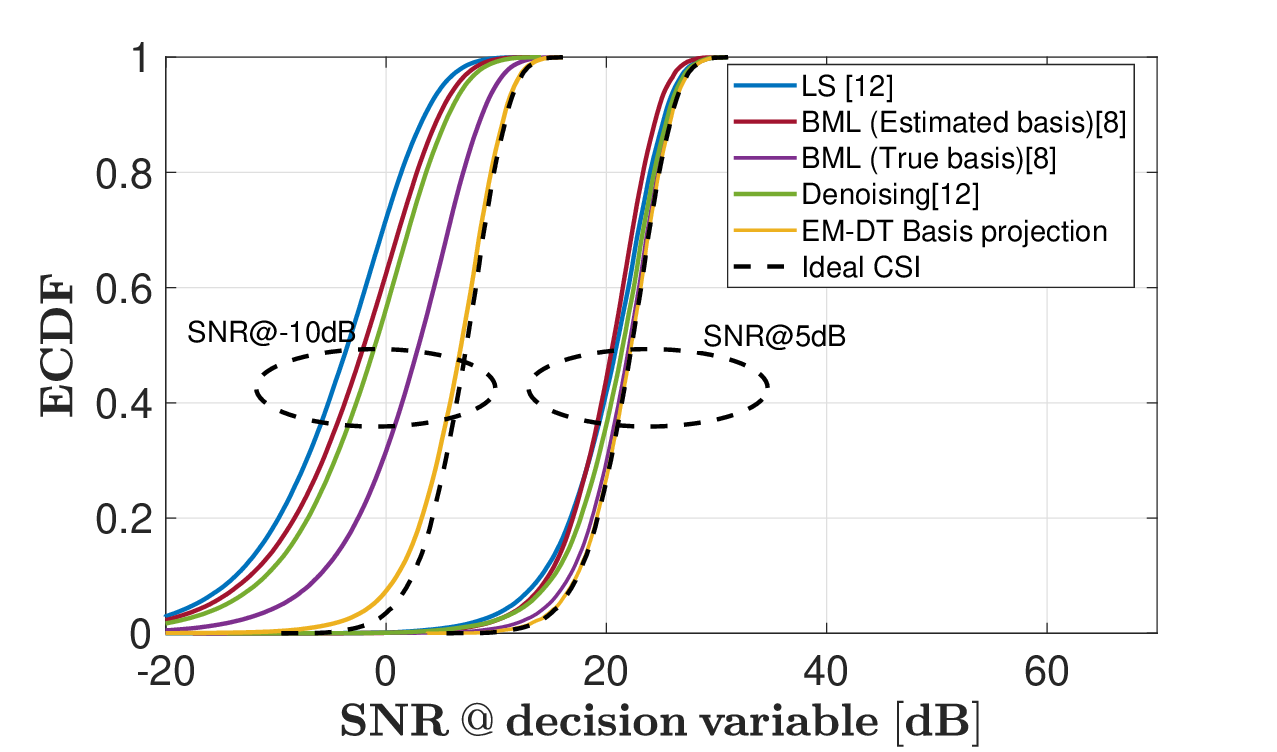}}
\caption{ECDF comparison for a SNR at the decision variable (after combing) when considering SNR $= -10$ dB and SNR $= 5$ dB among state-of-the-art approaches and our proposed method with $N_p=32$.}
\label{figECDF}
\end{figure}

Figure \ref{figECDF} confirms the improvement in spectral efficiency. In fact, evaluating the SNR at the decision variable, our proposed method allows to obtain higher performances with respect to the others in the low SNR region. 

\begin{figure}[!t]
    \centering
    \begin{subfigure}[b]{0.48\textwidth} 
        \centering
        \includegraphics[scale=0.44]{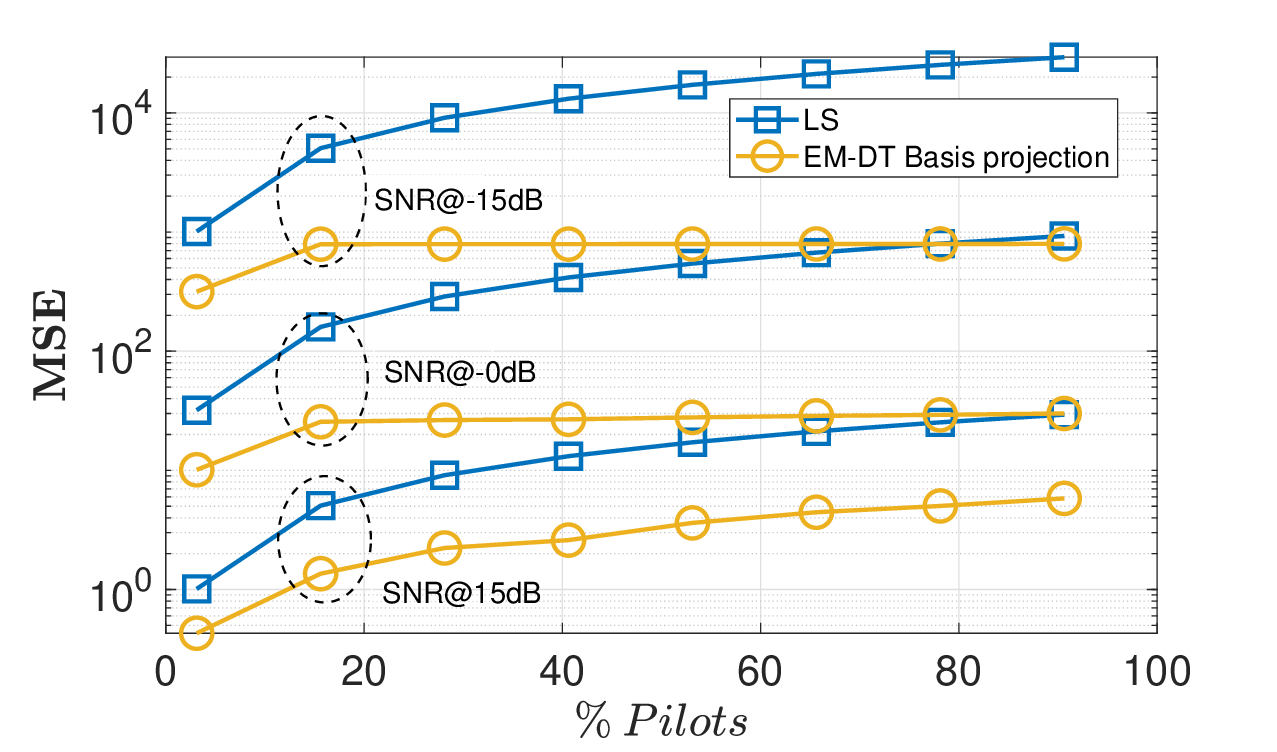}
        \caption{}
        \label{fig:subfig1}
    \end{subfigure}
    \hfill 
    \begin{subfigure}[b]{0.48\textwidth} 
        \centering
        \includegraphics[scale=0.44]{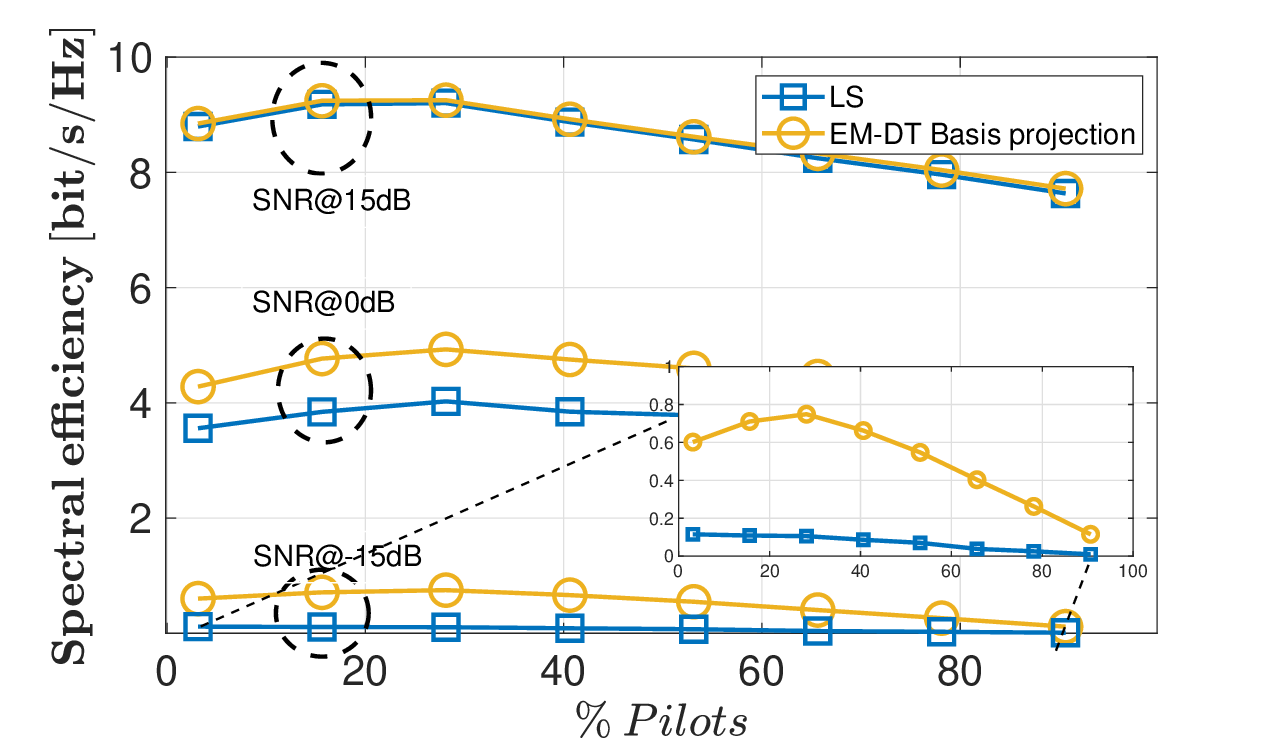}
        \caption{}
        \label{fig:subfig2}
    \end{subfigure}
    \caption{Comparison of a) MSE and b) spectral efficiency for the LS and the proposed approach, varying pilot subcarriers with respect to full piloting.}
    \label{figPILOT}
\end{figure}

Figures~\ref{fig:subfig1} and \ref{fig:subfig2} illustrate the MSE and the spectral efficiency over the pilot subcarriers for a SNR of $15\:\mathrm{dB}$, $0\:\mathrm{dB}$, and $-15\:\mathrm{dB}$, adopting $\mathrm{B}\approx 900\:\mathrm{MHz}$ ($N=2048$ subcarriers). As shown in \ref{fig:subfig1}, the proposed method shows a remarkable improvement in the MSE over the pilot subcarriers. In Fig. \ref{fig:subfig2},  both the LS and our proposed method increase until a peak in the spectral efficiency is reached, for $N_p=18$ (i.e. $28\%$ piloting). It is noteworthy that, at low SNR, our proposed method can drastically reduce the number of pilots, while maintaining a higher spectral efficiency in comparison with the LS approach. In fact, for $0\:\mathrm{dB}$ and $-15\:\mathrm{dB}$, our approach guarantees a higher spectral efficiency for $N_p=2$ (corresponding to $3\%$ piloting), with respect to the maximum value of the LS obtained for $N_p = 18$.

\section{Conclusion}
\label{Conclusion}
This paper proposes a new Bayesian channel estimation approach in the framework of EM-DT-empowered communications. Our proposed method exploits the a priori information constrained to a limited number of paths on delays and AoAs, provided by the EM-DT to perform a modal projection, with the aim of enhancing the channel estimation task. The new method is compared with baseline methods such as the LS, a BML and, a denoising approach. Simulation results show the improvement of performance in terms of NMSE and spectral efficiency with respect to current state-of-the-art methods. Specifically, our approach achieves a remarkable $20\:\mathrm{dB}$ improvement in NMSE compared to the LS method, along with a drastic reduction in the required pilots from $N_p=18$ to $N_p=2$. This reduction is achieved while maintaining superior spectral efficiency at low SNR with respect to the LS method.  Additionally, the error on the UEs position, which serve as input in the proposed framework to retrieve the spatial and temporal subspaces, should be investigated. 

\bibliographystyle{IEEEtran}
\bibliography{bibliography}

\end{document}